\documentclass[10pt,letterpaper]{article}
\usepackage[top=0.85in,left=1.5in,footskip=0.75in]{geometry}

\usepackage{changepage}

\usepackage[utf8]{inputenc}

\usepackage{textcomp,marvosym}

\usepackage{fixltx2e}

\usepackage{amsmath,amssymb}

\usepackage{cite}

\usepackage{nameref,hyperref}

\usepackage[right]{lineno}

\usepackage{microtype}
\DisableLigatures[f]{encoding = *, family = * }

\usepackage{rotating}

\usepackage{url}

\raggedright
\usepackage[aboveskip=1pt,labelfont=bf,labelsep=period,justification=raggedright,singlelinecheck=off]{caption}

\makeatletter
\renewcommand{\@biblabel}[1]{\quad#1.}
\makeatother

\date{}

\usepackage{lastpage,fancyhdr,graphicx}
\usepackage{epstopdf}
\fancyheadoffset[L]{2.25in}
\fancyfootoffset[L]{2.25in}

\begin{document}
\vspace*{0.35in}

\begin{flushleft}
{\Large
\textbf{Full Computational Reproducibility in Biological Science: Methods, Software and a Case Study in Protein Biology}
}
\newline
\\
Les Hatton\textsuperscript{1*}
Gregory Warr\textsuperscript{2**}
\\
\bigskip
\bf{1} Faculty of Science, Engineering and Computing, Kingston University, London, UK
\\
\bf{2} Medical University of South Carolina, Charleston, South Carolina, USA and Division of Molecular and Cellular Biosciences, National Science Foundation, Arlington, VA 22230.
\\
\bigskip

* lesh@oakcomp.co.uk
** gwarr@nsf.gov

\end{flushleft}

\section*{Abstract}
Independent computational reproducibility of scientific results is rapidly becoming of pivotal importance in scientific progress as computation itself plays a more and more central role in so many branches of science.  Historically, reproducibility has followed the familiar Popperian\cite{Popper1959} model whereby theory cannot be verified by scientific testing, it can only be falsified.  Ultimately, this implies that if an experiment cannot be reproduced independently to some satisfactory level of precision, its value is essentially unquantifiable; put brutally, it is impossible to determine its scientific value.  The burgeoning presence of software in most scientific work adds a new and particularly opaque layer of complexity\cite{Ince2012}.  In spite of much recent interest in many scientific areas, emphasis remains more on procedures, strictures and discussion\cite{Peng2011,Ince2012,Sandve2013,JoppaEtAl2013,NatureEditorial2014,NatureEditorial2014a,Freedman2015}, reflecting the inexperience of most scientific journals when it comes to software, rather than the details of how computational reproducibility is actually achieved, for which there appear to be relatively few guiding examples\cite{Claerbout1992,Donoho2009,Garijo2013}.  After considering basic principles, here we show how full computational reproducibility can be achieved in practice at every stage using a case study of a multi-gigabyte protein study on the open SwissProt protein database, from data download all the way to individual figure by figure reproduction as an exemplar for general scientific computation.


\section*{Principles}
This interdisciplinary paper falls squarely between two juggernauts: the sciences which consume computation as a necessary tool and the study of computation itself.  We hope it successfully bridges both to provide a research scientist with enough information on exactly how to achieve reproducibility in practice and provides a computer scientist with some context in how important the simplest of tools can become.

We will first try to present typical viewpoints.  To a scientific user of software, software can be an annoyance - it is frustrating, time-consuming and essential but it is ``only'' programming and therefore not particularly difficult compared to the real science.  As such it is generally under-resourced in scientific work and is often left up to hard-pressed research assistants to do in their spare time in spite of being generally untrained in computational tools and techniques.

On the other hand, to a computer scientist, reproducibility is merely an interesting deployment of existing tools and techniques - but not particularly novel.  \textit{In contrast actually guaranteeing software is free of error is not only novel but fundamentally intractable in the sense that nobody knows how to do it, or even if it can be done.}  As such, verifying even a modest sized program as correct in function is far harder than doing the science in the first place.  This remains an unpleasant truth in spite of the extraordinary progress of Computer Science in the last fifty years or so, and CS researchers still find the elimination of defects from programs and the quantification of the results of a scientific computation challenging to say the least.  Quoted residual error rates for software, that is error rates \textit{after} it has been released, remain stubbornly high and may be anywhere from around 0.1 per kSLOC (thousand lines of code) for the very best \cite{Kell93} to 100 times worse, \cite{BasiliPerricone1984,Hat97b}.  The low end of these figures represent impressively reliable systems but the rapid growth in size of software exemplified by a few of the examples in Table \ref{table:sizes} suggests that there are still likely to be a significant number of residual and generally unquantifiable defects, even when packages are open source and as widely inspected as these.  The code written by researchers is likely to be rather worse than this for no other reason than that insufficient time and resources is available to get them this good.    Even public submissions to prestigious algorithm collections such as CALGO, where it might be expected that a higher degree of preparation would be present, are often found wanting, with various repeating defects and missing tests, \cite{Hopkins2002a}.

\begin{table}[ht]
\centering
\begin{tabular}{ccccc}
\hline
Package & Language & kSLOC \\
gimp-2.6.11 & C & 902.7 \\
perl-5.14.0 & C & 270.0 \\
mysql-5.5.13 & C & 773.5 \\
R-2.13.0 & C & 433.9 \\
gcc-4.6.2 & C & 3761.8 \\
gnat-3.15p-src & Ada & 575.7 \\
kdebase-runtime-4.6.3 & C++ & 170.4 \\
kdelibs-4.6.3 & C++ & 1291.9 \\
lapack-3.3.1 & Fortran & 953.9 \\
linux-2.6.39 & C & 13263.6 \\
openjdk-7-ea-src-b143-20\_may\_2011 & Java & 4289.4 \\
\hline \\
\end{tabular} 
\caption{Typical sizes of packages in wide use in scientific research along with the revision numbers \cite{HatTSE14}.}
\label{table:sizes}
\end{table}

It is probably worth noting that if a researcher ever managed to produce an error-free program of any significant size, a) they would not know it because they would be unable to prove it, b) they would not live long enough to be able to test it completely and c) they would be unable to repeat the process systematically.

Indeed it is a common misconception to think of a program as \textit{tested}.  Unfortunately, it is difficult even to define this concept in CS

\begin{itemize}
 \item Tested under what conditions ?
 \item What were the inputs ?
 \item What were the outputs ?
 \item How many of the possible paths in the program were exercised by the tests ?
\end{itemize}

With regard to the last of these, Adams \cite{Adm84} in a celebrated study, demonstrated on IBM's then world-wide mainframe distribution that about a third of all the faults found took \textit{longer than 5,000 execution years} to manifest themselves, greatly undermining any possible role for dynamic testing as a means of veryifying code.  This means that a fault occurring this rarely would be seen approximately once in 50 years in just one of 100 machines operating independently.  For a recent review, see Hatton \cite{Hatton2011a}.

In parallel with the growth of ancillary packages such as those in Table \ref{table:sizes}, researchers write more and more software in their own projects, mostly written in programming languages which have been in use for decades, \cite{Rousseau2012,Holzmann2013}.  These latter references also give some important insights into modern methods of eliminating as many defects as possible in scientific environments.  However, such methods cost a lot of money and time, and hard-pressed researchers rarely have the budget, technical interest and resources, and probably not the know-how to perform such packaging and verification.  It would be tantamount to asking a particle physicist to do their own brain surgery.

\subsection*{Code sharing}
Unfortunately, premier scientific journals continue to conflate simple \textit{code sharing} with full computational reproducibility even in influential editorials \cite{NatureEditorial2014}.  Even on the question of code sharing the language is equivocal, and almost apologetic  - \textit{``Nature journals have decided that, given the diversity of practices in the disciplines we cover, we cannot insist on sharing computer code in all cases''}.  This policy makes clear that even (or perhaps especially) elite journals are quite flexible when it comes to accepting degraded standards of scientific practice in certain fields whose results they wish to publish.

We understand the logistical problems but code sharing alone is a relatively small part of full computational reproducibility and it is simply not correct to think otherwise.  To see the difference between simple code sharing and providing the complete means to build and run code, we have only to look at the world of software escrow \cite{Binder2013} whereby commercial software source code is held by a third party as part of a commercial agreement.  The intent of an escrow agreement is to give an  end-user access to the source code of a proprietary system in very special circumstances controlled by the escrow agent, for example if a software supplier becomes insolvent.  Such escrow agreements must not only include the source code but also the complete means to build the system in a form which the end-user can use and to satisfy a hopefully rigorous set of test procedures to give confidence that it is functioning appropriately.  This is a non-trivial step and it is a sobering thought that the computational reproducibility study which we exemplify here used a Linux machine, the GNU compiler, perl, R and additional other packages so is built on top of at least 20 million lines of code.  Printed out at 100 lines of code per page, it would produce a stack of paper around 20m high.

\subsection*{Virtual machines}
This is worth mentioning in that putting up results using a virtual machine in the cloud somewhere is a possible way of distributing a scientific result.  Unfortunately, it is effectively useless as a mechanism for establishing computational reproducibility without appropriate packaging.  If the machine is simply a binary engine which will produce the results when a third party presses the button, then it is not \textit{reproducibility} but simply \textit{replicability} about which we will have more to say below, and tells you nothing about the nature of the computations which led to the result.  If the machine is an image of the researchers machine without any of the appropriate packaging described below, it is simply code sharing and remains unquantifiable.  In short, a virtual machine is only a convenient vehicle for delivering computational reproducibility; it still must follow the processes described in what follows.

\subsection*{Nomenclature}
There are also problems with nomenclature.  The terms code or computercode\cite{NatureEditorial2014} are simply too ambiguous to be useful.  The \textit{binary code} (machine-readable) is not useful because it is unquantifiable.  The \textit{source code} (human-readable) is a minimum requirement but quite apart from the difficulties described above, there are other significant drawbacks\cite{Ince2012}.  We define \textit{full computational reproducibility} as a set of open deliverables which provide the \textit{complete} means to reproduce all of the published results.  As such the reproducibility deliverables should not only include all source code, but should also include all ancillary source code necessary to build and run the analysis programs, ancillary source code to download the datasets (unless they are included as part of the package), ancillary source code scripts to run statistical or other analyses and \textit{finally a regression suite produced by the original research team against which any locally reproduced copies will be tested automatically}.  Anything less than this surely fails to meet the definition of rigorous science; the reader is without the means to verify, even at the simplest level, the validity of the conclusions.

Full computational reproducibility is not difficult in principle, particularly with the modern availability of open source tools and data, but it is very technical and time-consuming work. Although much can be done to streamline the process, it is clear that many practicing scientists still either question or are unaware of its importance.  We emphasize that full computational reproducibility is a vital extension of the scientific method as espoused by Karl Popper \cite{Popper1959}.  Researchers have long noted the importance of Popper's work in various contexts \cite{Ziolkowski1982} but in a scientific age increasingly reliant  (and in many fields completely dependent) on computation, full computational reproducibility has, we would argue, become a \textit{sine qua non}.  Without the implementation of full computational reproducibility, it will be impossible to determine whether, and to what degree, any computationally-dependent results have been corrupted. Unfortunately the reader has no way to determine which results in the published literature are correct and which are not.  The true costs of non-reproducibility are now beginning to surface\cite{Freedman2015} and appear to be high indeed; thus any initiatives to improve reproducibility\cite{ScienceEditorial2014,NatureEditorial2014a} are laudable.  Unfortunately, statements such as ``... just because it is not reproducible does not necessarily make it wrong.'' \cite{ScienceEditorial2014} are remarkably rash and unhelpful from a premier journal.

That there is serious problem is highlighted in a recent news item in Nature\cite{Nature2016}, where 52\% of the respondents thought there was a 'significant crisis', however of the factors cited for this crisis by respondents, software problems rated only a sub-category under methods.  When asked which factors could boost reproducibility, software transparency was not even mentioned.  This represents a woefully large chasm between the providers of software methods, the computer scientists, and the consumers of software methods, the research scientists.  As the news item stated, it is perfectly true that statistical techniques for example are frequently inadequately or even incorrectly applied, however, this is a failure of peer review.  We have had decades of careful statistical science to guide such methods.  In contrast, peer review of software is in its infancy to say the least.  Software methods are indeed in danger of becoming the ``elephant in the room'' in discussions of scientific reproducibility.

Before discussing the general principles of computational reproducibility, it is worth emphasizing the following point:

\begin{quotation}
\textit{Full computational reproducibility is the very MINIMUM requirement for a computationally dependent result to qualify as rigorous science.  It does not guarantee that results are correct but it does guarantee that not only can they be repeated to an acceptable level of significance, but that every step of the process is open for independent checking.  This is exactly analogous to conventional experimental reproducibility, the cornerstone of scientific progress for many years.}
\end{quotation}

Of course, even if the complete means for computational reproducibility is available, there is no guarantee that anyone will bother to examine the study \cite{Easterbrook2014}.  However, that should not be taken as an excuse and if it is made as simple as possible, the chances of independent reproduction of a computationally dependent result are larger.

\section*{Computation in science}
\begin{figure}
\centering
\includegraphics[width=12cm]{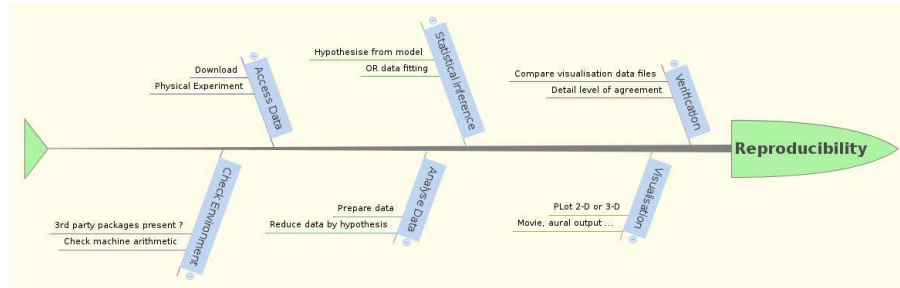}
\caption{The necessary elements to achieve full computational reproducibility in a typical scientific experiment.}
\label{fig:reproduce}
\end{figure}

We will discuss the general role of computation in an archetypal scientific experiment using Figure \ref{fig:reproduce} where the contributing elements in enabling full computational reproducibility are shown from the left.  We discuss each of these from a general point of view here and will later specifically attach openly available pieces of software to them with a discussion of how they work before showing finally, the results of them working.

\begin{description}
 \item[Check computational environment] The computer environment consists of the set of programs which the researcher may use to produce their results and the hardware on which they run.  These might be stand-alone packages such as \textit{gnuplot} (plotting data) or \textit{R} (statistical inference), interpreters such as \textit{perl} or \textit{python} which run the researcher's own or other perl or python programs, or compilers such as \textit{gcc}, the GNU compiler for numerous scientific languages including C, C++, Fortran and Ada.  An important aspect is to make sure that the computer does arithmetic correctly to the internationally standardised IEEE 754 \cite{Beebe2007}.
 \item[Access Data] All scientific experiments are underpinned by data that can often be downloaded as is the case with the protein case study we describe shortly to exemplify these principles.  It may equally be data acquired by the researcher using lab equipment, although to add a further layer of complexity, a researcher might be astonished to find how much software is actually embedded in the lab equipment itself, which is not without its own problems \cite{Genuchten2007}.  This proprietary software is generally unquantifiable and has to be taken on trust at present although we will expand on this later.
\item[Analyse data] Here the raw data is processed in various ways to isolate subsets of importance.  This is particularly so in modern science where datasets may be immense.  The case study we use to exemplify reproducibility utilises a raw dataset of around 3Gb, (taken as $3 \times 10^{9}$ bytes). 
 \item[Statistical Inference] In all scientific experiments, a measure of statistical significance is essential to match theory with observation.  This process might use \textit{R}, \textit{SPSS} or less commonly, hand-crafted software for example.
 \item[Visualisation] As a glance through almost any recently published journal will reveal, articles increasingly contain (especially in the Supplementary Materials) extremely large datasets and complex illustrations.  These may have been produced in numerous ways and may include complex 3-D representations, animations, sound recordings and so on.
 \item[Regression] This is truly fundamental and provides a simple way for researchers trying to reproduce the results of a scientific paper to measure objectively how close they are.  Simple inspection of graphical outputs, however exotic, is insufficient \cite{HatRob94}.
\end{description}

\section*{Some comments on reproducibility}
The object is to integrate as closely as possible the stringent reproducibility recommendations of Ince et. al \cite{Ince2012} with the above model to allow the full reproduction of all the computational results of a paper for independent corroboration.  In this sense we will go much further than detailing guidelines for good practice \cite{Sandve2013} by providing an actual example, although this naturally embodies a number of these guidelines.  The model and examples will we hope be adaptable by others for their own experiments.

We are not the first to attempt this.  Other researchers have produced similar initiatives, \cite{Claerbout1992,Donoho2009}.  Notably, Jon Claerbout has been grappling with this thorny problem for 30 years with his group at Stanford ``an article about computational results is advertising, not scholarship. The actual scholarship is the full software environment, code and data, that produced the result.'' \cite{Claerbout1992}, but the practice is as yet by no means widespread in the face of the now formidable levels of computation in some fields.  More recently Dave Donoho published an invitation in the journal Biostatistics to researchers \cite{Donoho2010} so there is some momentum growing but relatively slowly given the general reluctance of some journals to grasp this particular nettle.

We will distinguish the following types of principal computational object which might be used by a researcher in their work, (there are others but these give useful definitions for our discussion).

\begin{description}
 \item[Specific programs and configuration]  These include everything written by the researchers in either compiled languages such as Fortran, C, C++ and scripted languages such as \textit{perl} or \textit{python}.  Any configuration files necessary to run these hand-crafted programs will also have been written by the researchers.  In addition, the environmental files such as the \textit{make} scripts to allow them to be made locally would also be included.
 \item[Third-party programs and configuration] These include programs readily available to the researcher but not written by them.  These would include applications such as \textit{perl}, the statistics program \textit{R}, the plotting program \textit{gnuplot} or whatever.
 \item[Operating System and configuration] This is the operating system on top of which the specific programs and the packaged programs would be run.  It might be virtualised, proprietary such as Windows or OS X or open source such as Linux or BSD.
\end{description}

The essence of computational reproducibility is to enable a reasonably knowledgeable researcher in the same field to set up a computer without too much effort using a reproducibility package supplied by the original researchers in such a way that the original computational results can be reproduced to some reasonable level of accuracy, which should be stated, just as if they were repeating a physical experiment.  This should consist of all source code actually written by the original researchers and as much as practicable of the rest as is discussed below.  It should NOT simply be copies of the original binary executables.  The output of these is not amenable to independent verification unless the source code is available - neither binaries nor an algorithmic description are sufficient, \cite{Ince2012}.  In the case of open source, it would be sufficient to quote the version number of packaged programs. for example, the GNU compiler \textit{gcc}, or the \textit{perl} environment in use, because the original source code of most such releases is readily accessible.

It may be that source code of third-party programs is not available.  In that case, the ability to reproduce the results is degraded and must be replaced by trust in such programs.  We would argue that in view of the preceding discussion, it is naive to trust any computer program as being 'correct', and in an ideal world, envisage the source code of all programs to be continually revisited to reduce their errors and increase confidence in their output incrementally inline with the scientific method.  This employs a process known as source code inspection, the efficacy of which has been understood for almost 40 years, \cite{Fag76,Hum95,Koru2008}.  Clearly, the use of open source has revolutionised the possibility of reproducing computational results and this opportunity should be grasped with both hands.  Without this, and to repeat our earlier perhaps rather bleak conclusion, \textit{if the software used in a scientific experiment is not independently verifiable, then the results fall outside the definition of rigorous science and their value is unknowable.}

The experiment described here was carried out entirely with open source and is therefore theoretically reproducible at the highest possible level since everything can be inspected if necessary, and reproduced to the limits of single-precision floating point arithmetic, since the arithmetic on this machine is IEEE P754 conformant as noted shortly.

This protocol is not intended to be prescriptive in any way, it simply details what was done to allow the results to be fully reproducible, thus we hope that it approaches the ``Gold standard'' described by Peng \cite{Peng2011} and gives a useful guide to other researchers as the principles are the same even if their software environment, data flow and analysis are somewhat different.

\subsection*{Software environment}
Whilst it is possible to run this software in a Windows environment, this entails the installation of a number of open source packages \cite{Cygwin2015,DGJPP2015} and their execution on a system which is proprietary and therefore closed.  This in no way implies that Windows is an inappropriate environment.  It simply makes the point that Windows of itself is not independently assessable for incorrectness.  Linux on the other hand is open source and \textbf{is} at least in theory, independently verifiable allowing us to claim that every piece of software used in the production of the scientific results described here is capable of independent open scrutiny.  This does not imply that either is more reliable than the other, simply that Linux is more independently accessible.

We don't for a moment imagine that any reader of this is going to download the 14 or so million lines of code in the current Linux kernel accompanied by several million assorted lines of code in the packaged programs which were used such as \textit{perl}, the \textit{bash} shell and the redoubtable GNU \textit{gcc} compiler and check through them before embarking on their project, but the point is that they could.  In any case, these pieces of software are independently scrutinised every day by many application experts.  Equally, it would not be expected that the researchers at CERN would take apart every single component of the LHC before running an experiment.  The scientific method does not demand perfection.  It simply demands that we collectively might continue incrementally verifying layer upon layer until something is beyond reasonable doubt and that nothing should ever impede this process.  Software has greatly complicated this accessibility by reason of its rapidly growing size and complexity but open source at least guarantees that it can if necessary be done.  This seems to be on the right track in the long term  but the necessity for researchers to release in a useable form, the source code they themselves wrote, remains paramount.

\subsection*{Third-party packages}
All modern systems allow the version number of their components to be identified.  For example, Linux has simple although rather variable ways of detailing system versions as revealed by the following, detailing the principle third-party packages which were used, (commands appearing in bold) and their output.

\begin{description}
 \item[more /etc/*-release] PRETTY\_NAME="openSUSE 12.3 (Dartmouth) (i586)"
 \item[uname -a] Linux linux-qa7l.site 3.7.10-1.45-desktop1 SMP PREEMPT Tue Dec 16 20:27:58 UTC 2014 (4c885a1) i686 i686 i386 GNU/Linux
 \item[perl -v] This is perl 5, version 16, subversion 2 (v5.16.2) built for i586-linux-thread-multi
 \item[gcc -v] gcc version 4.7.2 20130108 [gcc-4\_7-branch revision 195012] (SUSE Linux)
 \item[gnuplot --version] gnuplot 4.6 patchlevel 1
 \item[R --version] R version 3.1.2 (2014-10-31) -- "Pumpkin Helmet".  (We also used the third-party R packages \textit{tgp}, \textit{akima} and \textit{plot3D} which may be installed within the R environment using the command \textit{install.packages("tgp")}, and so on.)
\end{description}

This information is sufficient to take a standard PC and install it to the same specification as was used in our study.  This is not an onerous activity and should be mandatory as the essence of reproducibility is to provide the means to reproduce results.  Whether or not a researcher chooses to go to such lengths is up to them, however for a really pivotal result, there is no reason not to do it and relevant budget and resources should be part of the overall research study, \cite{Ince2012}.

\subsection*{Arithmetic verifiability}
Although it should not really be necessary in this day and age, it is always comforting to verify that the chipset / compiler combination actually does arithmetic reasonably correctly.  Many scientists will take this for granted but sufficient implementation errors have appeared in computational units over the years, from the notorious FDIV Pentium bug onwards, (http://en.wikipedia.org/wiki/Pentium\_FDIV\_bug), quite apart from implementation deficiencies in languages such as Java as to make it worthwhile\cite{Kahan2004,Kahan2011}.  In addition, compilers get upgraded from time to time, perhaps not always correctly and it is always worth checking as it only takes a moment.  There is an international standard for floating point arithmetic (IEEE 754) and numerous freely available test programs\cite{Beebe2007}.

As all of the above named programs, perl, R, gnuplot and so on in common with many open source packages used by scientists are written in the programming language C, its intrinsic model of arithmetic computation is the bedrock of much open source computation.  An open source method for verifying arithmetic performance in both standard and embedded systems is \textit{paranoia}, which only requires an ISO compliant C compiler and for which \textit{gcc} serves admirably \cite{Kahan1986,Hatton2005}.

Running it on the environment used in the accompanying paper \cite{HattonWarr2015} gave in summary:-

\begin{verbatim}
     Embedded System Paranoia SUMMARY
         SINGLE PRECISION 32 bits
 Closest separation = 5.9604645e-08

 Number of FAILUREs encountered       = 0
 Number of SERIOUS DEFECTs discovered = 0
 Number of DEFECTs discovered         = 0
 Number of FLAWs discovered           = 0

PASSED : No failures, defects nor flaws have been discovered.
 Rounding appears to conform to IEEE standard P754.
PASSED : The arithmetic diagnosed seems Excellent.

 Rating ...

         =====> Excellent
                Very good
                Good
                Acceptable
                Unacceptable
                Broken
...
\end{verbatim}

This speaks for itself.

Note that this is not quite as trivial as it might appear even though very little C was actually written for the case study we describe.  However, the main application packages, notably \textit{perl} and \textit{R} used here are themselves written in C and use the arithmetic model of C.  This verification step is therefore more important than it might first appear as it is the arithmetic bedrock on which a very large amount of practical computation sits.

The position of C as a computational \textit{lingua franca} used to program the vast majority of open source projects, (probably greater than 75\% of all code \cite{HatTSE14}), means that such verification is very likely relevant in many projects.

\subsection*{Data formats}
In analyses of large data, it proves convenient to keep all intermediate files in some kind of human-readable form.  Conversion costs between binary and human readable form can generally be ignored and proved completely negligible on this project and human readable means that:

\begin{itemize}
 \item Testing is much simpler
 \item Many modern data analysis tools such as \textit{perl} and \textit{grep} work directly on human readable formats.
 \item Human readable formats compress very well using \textit{zip}, \textit{gunzip} or an equivalent if storage is at a premium.
\end{itemize}

With regard to data formats, the need for human readability goes far beyond just the raw data.  Any proprietary file format should be considered anathema because they do not have one of the most important requirements for reproducibility - longevity.  To illustrate, one of us (LH) has examples of C code written to ISO C 9899:1990 which is over 35 years old.  The source code files (in ASCII) can today not only be read by any of a number of editors, closed source or open source, but it will compile and run with current versions of the GNU C compiler and give the same answers as it did over 35 years ago.  In stark contrast, the same author has examples of technical papers written in an old but proprietary version of Microsoft Word for Apple less than 15 years ago which cannot be read with any current version of Word on Apple or PC or indeed anything else.  The only thing that can be retrieved is the text - everything else, figures, equations and formatting is lost forever.

Journals which insist on Word submissions should take note.  In contrast, Latex files are human readable and the Latex system itself is open source.  Not only that, but as explained later, they can be used as the basis of a system which can not only reproduce the scientific results, but reproduce the final published format of a journal article, all from the same file, including equations, tables, references and figures.

\subsection*{Specific programs and source code control}
Researchers have many implementation languages from which to choose.  Which is the best of the hundreds available ?  This question has been teasing computer scientists for decades and there is no simple answer.  Choice breaks down roughly to:-

\begin{description}
 \item[Compiled languages] These include the juggernauts of the programming world, \textit{C}, \textit{C++}, \textit{Fortran}, \textit{Ada} and so on.  Writing in these languages requires specialist knowledge of the myriads of ways of getting it wrong, \cite{SaferC,HatMISRA1}.  They can be very highly portable and thanks to the GNU C compiler, ubiquitous.  Fortran in its most recent incarnations F90, F95 is particularly good at dealing with vector and matrix quantities and has a fairly sane arithmetic environment.

Although there are libraries available for these languages, the situation is nowhere near as good as it is for the next category.
 \item[Interpreted languages] These include languages like perl and semi-interpreted languages like Java.  Perl, like, C++ and so on, is old but it has borne the test of time well.  There are myriads of freely available libraries, most of them with inbuilt test suites, and it is probably unmatched as a pattern recognition language.  When the data is in human readable form, perl is quite literally invaluable.  Java is harder work, derived as it is from C and C++.  The interpreted languages as a whole have attractive properties when compared with the compiled languages,  \cite{Prechelt99}
 \item[Hybrid systems]  It is not difficult to produce systems which are heterogeneous in the sense of having different languages for different parts of the system.  These can communicate loosely through shared files or be more tightly integrated.
\end{description}

Although language choice invites often passionate responses, the authors' view is that it doesn't really make much difference.  It's usually better to pick something you are already good at and its helpful to think of the act of choosing programming languages as a fashion statement rather than anything much to do with engineering.

Source code control is of absolutely vital importance so that which revision of software actually in use (and there will typically be many even on a simple research project) can be readily identified.  Developers will generally find that little and often (known as incremental development) is much more controllable than big bang (loads of new features, switch on, step back and see what happens).  For  local systems, revision control using the Linux utility \textit{rcs} is very robust.  For bigger and more distributed teams, \textit{git} or \textit{subversion} is highly recommended.

\subsection*{Packaging for reproducibility}
Packaging for reproducibility is a painstaking and laborious process.  Even when the workflow in a study is carefully documented, it may take a very considerable amount of time to reproduce a result \cite{Garijo2013}.

It is particularly difficult to do when the research is continuing because computationally based research into text analysis as is the case with protein databases is about tool-building.  To investigate a potential data pattern, a \textit{perl} tool can be assembled in a couple of hours without performance fears and used to explore the data.  There is therefore an overwhelming tendency to rush ahead leaving streams of small and often discarded tools behind.  Even computer scientists do this.  In stark contrast, to facilitate independent verification, reproducibility is a careful reconstruction of the simplest and most comprehensible path to reproducing the claims, which are typically figures or tables of data analysis.

In our case, the original working directory for our project contained 5,679 lines of \textit{perl}, shell and \textit{R} scripts, C programs and makefiles.  The packaging process distilled this down to 3,812 lines measured the same way.  This process of rationalisation is essential because research inevitably builds up clutter whilst ideas are pursued, especially software clutter because it is so easy to write, and although the whole directory could be simply bundled up and released as reproducible, in practice it would not be because it is relatively unstructured.

We re-iterate that \textit{the essential purpose of reproducibility after all, is not to tick boxes so that it is merely possible, but instead to encourage people to actually do it, improving the quality of all computational science as projects build one upon another.}  Attempting to make sense of 5,679 lines of relatively unstructured research code is not for the faint-hearted and does not represent a reproducibility deliverable.  Research code can be much larger than this and results must be as easy to reproduce as the originating research team can make them.

On another point, it was not immediately obvious but efficiency turns out to be very important to this process because assembling the packaging means a seemingly endless repetition of clean installs and incremental testing to make sure that the figures and tables are faithfully reproduced and, not insignificantly, still agree with the claims in the original paper !  Unless these can be run quickly and efficiently, it is too easy to cut corners.  In our case, it took a full three days of running scripts repeatedly, streamlining and checking and re-running before the packaging was complete.  As the entire suite now runs in less than a minute on a reasonable desktop, there were literally hundreds of re-runs as the packaging was shaped into its downloadable form.

Naturally, this effort takes a lot of time and we would urge, again following \cite{Ince2012}, that in any funded research, budget be allocated for this or the value of the science will be greatly undermined.  Even for a practising computer scientist, the overall exercise might reveal surprises.

\section*{Full Computational Reproducibility: a case study}
We will now put flesh on these concepts by demonstrating how it can be achieved in practice in a computationally intensive research study of the SwissProt database, \cite{HattonWarr2015}.  The resulting reproducibility package is freely downloadable from http://leshatton.org/\footnote{see http://www.leshatton.org/pone\_0125663\_reproducibility.html}.  In an ideal world, ``Ideally, an external researcher should be able to download a reproducible research work and reproduce all the results with a simple mouse click'' \cite{Vandewalle2009}.  In theory, we could have done this but it seems more appropriate to a researcher to be able to home in on a particular Figure or Table as well as being able to generate all the Figures and Tables.  Doing this from the command line might be considered ``nerdy'' \cite{Donoho2010}, but it is no more nerdy than addressing an envelope and it has the benefit of complete transparency.  Introducing some kind of graphical user interface with its associated libraries on top simply adds another large and potentially error prone layer of software, (around 1.5 million lines of C++ in the case of KDE, Table \ref{table:sizes}).

\subsection*{Deliverability package download and environmental work}
First the reproducibility package itself is downloaded as a zip file onto a suitable Linux platform and unzipped in a convenient directory. The package is organised into the following phases with a preceding digit to indicate the phase at which it should be run, starting with 0 (0- indicates environmental checking, 1- data download, 2- data preparation, 3- figure and table reproduction, 4- verification).  They should be run in this increasing numerical order and we will specifically reference the components here, comparing them directly against Figure \ref{fig:reproduce} from the left.

\subsection*{Check environment}
\paragraph*{Command-line: ./0-environment.sh}

This command does two things.
\begin{itemize}
 \item 3rd party package checking.  Using a series of statements like the following:-
\begin{verbatim}
which gnuplot       2> /dev/null
if test $? -ne 0
then
     echo "gnuplot MISSING"
     status=1
fi
\end{verbatim} 
This simply uses the Linux command \textit{which} to see if the package \textit{gnuplot} is present and report it if not.  This is repeated for all the 3rd party packages actually used in the study.
 \item Check machine arithmetic. After verifying that all the necessary packages are present, the script then makes the built-in copy of \textit{paranoia} and runs it.  The report on the machine arithmetic is produced in the file \textit{esp.out}.
\end{itemize}

If all the necessary components are present and the machine passes \textit{paranoia} acceptably, the experiment is ready to be reproduced.

\subsection*{Access data}
This phase loads the third-party datasets.  Such datasets might be generated locally but in this case, these were downloaded from the standard SwissProt repository at http://uniprot.org/ and the Selene data repository for Post-Translationally Modified (PTM) amino acids at http://selene.princeton.edu/PTMCuration.

\paragraph*{Command-line: ./1-downloadUniProt.sh}
This carries out the following. 

\begin{itemize}
 \item First it downloads the complete current distribution using the following command.
\begin{quotation}
 \textit{wget} \url{ftp://ftp.uniprot.org/pub/databases/uniprot/current_release/knowledgebase/complete/uniprot_sprot.dat.gz}
\end{quotation}
 \item It then decompresses it using \textit{gunzip} leaving a raw data set \textit{uniprot\_sprot.dat} of some 2.7Gb.  This dataset follows best practice by being human readable in a format which is relatively easy to parse using simple \textit{perl} programs.
 \item Finally, it unpacks it into kingdoms and species using a locally produced \textit{perl} program, \textbf{uniprotSplit.pl}.  This program is less than 100 lines of perl and splits the raw data set using the built-in ID and OC headers.  In addition instructions are given on how to download the much smaller Selene PTM dataset for merging with the SwissProt dataset.
\end{itemize}

\subsection*{Analyse data}
This is a relatively complicated part of most scientific experiments.  The data has to be carefully organised, reduced if possible for efficiency and then analysed.  In Figure \ref{fig:reproduce}, we describe this in two stages.

\begin{itemize}
 \item Prepare data. At 2.7Gb, SwissProt is a fairly large raw dataset given that it would be processed thousands of times in our researches, so some simple experiments were carried out to pack the data into a form which would take advantage of pipelining in modern chip sets.  The first point to note is that instead of feeding individual species files in one by one in .fasta format, (which contains much extraneous information for our needs), they would be slimmed down by parsing out extraneous information, reducing the parts that were needed to a common .csv (comma separated variable) format with one record per protein.  This is illustrated below.

\begin{verbatim}
for file in *.fasta
do
   cat $file | processing ...
done
\end{verbatim} 

is \textbf{much} less efficient than

\begin{verbatim}
cat $combined_file | processing ... sieving as necessary
\end{verbatim} 

Here \textit{combined\_file} is produced just once and then repeatedly analysed.  It is worth spending a little time at the start of a research project doing this because it really pays dividends later.

The .csv format is a particularly easy text format to work with and there are many support tools available.  In addition, fields were added within this record to identify the kingdom, species and protein name so individual groupings could be trivially extracted using simple filters like \textit{grep}.  During this stage, the independent Selene information was also merged into the .csv records for each protein.  The data could have been loaded into a MySQL database or similar but csv format is perfectly suitable and very much more efficient on this scale.

A second stage of data preparation concatenated all the files into one.  From now on, every analysis stage started by piping this single file into a chain of other filters as above.  This reduced a single processing stage from around 30 minutes to only seconds on our systems.  Repeating the same processing thousands of times while exploring the dataset is then entirely feasible.  We refer to this as \textit{data rummaging} as being similar to data mining but when you don't really know what you are looking for as is often the case in the early days of a research study.
\item Reduce data by hypothesis.  By this we mean extracting those parts of the dataset which are identified by any underlying theory as most relevant.  When datasets are large and ours was quite large, this is an important step.  In our study \cite{HattonWarr2015}, our theory predicts that in proteins, the total length of the $i^{th}$ protein in amino acids, $t_{i}$ and its unique number of amino acids including any post-translationally modified amino acids $a_{i}$, formed a pair of canonical variables related by

\begin{equation}
t_{i} \sim a_{i}^{-\beta}
\end{equation} 

To confirm this prediction, the data was reduced on these variables using the locally produced \textit{perl} program \textbf{txtToTok.pl}.  There is some other infrastructure of shell scripts involved in this as shown in the reproducibility package but the bottom line is to produce a reduced canonical dataset from which Figures and Tables can be produced which explore how well this relationship is satisified by the data.
\end{itemize}

This is all carried out by the following script which also has an argument N.

\paragraph*{Command-line: ./2-prepareData.sh N}

The value of N considers the three different perturbations to the generated tokens to probe the robustness of our results are described in our paper \cite{HattonWarr2015} relating to the counting of Methionine start codons in the proteins, the excision of signal peptides and a Monte-Carlo estimate of unique amino acid alphabet counting.  These can each be obtained by modifying the value of N as follows:

\begin{description}
 \item[0] The data is processed without adulteration, in other words as it appears in the SwissProt release.
 \item[1] The data is processed by normalising all Methionine start codons
 \item[2] The data is normalised by excising all signal peptides.
 \item[3] The data is normalised by applying Monte-Carlo methods to the unique alphabet counting of the otherwise unadulterated data.
\end{description}

The net effect of this data preparation phase and canonical reduction is to reduce the original 2.7Gb. dataset to a far more manageable one of some 76Mb., a reduction factor of almost 36.

\subsection*{Statistical inference and visualisation}
We separated these in Figure \ref{fig:reproduce} but here it is convenient to merge them into one stage as much of this was done with the excellent open source \textit{R} program, \url{http://r-project.org}.

Ideally, the final paper and the results should both be reproduced from the same file.  This is quite simple to achieve in Latex by embedding some recognisable unique pattern in the .tex source file which annotates the reproducibility for the reader but also can be pre-processed to generate the set of executables which actually then generate all the figures and tables.

One way of doing this is to use a Latex macro, for example:

\begin{verbatim}
\newcommand{\reproduce}[2]{{\it The reproducibility script is \bf #1 #2}}
\end{verbatim}

In a table this might be referenced in the table or figure caption, or perhaps in the text after the first occurrence in those journals which require figure captions to be separated from the main text, as

\begin{verbatim}
... \reproduce{./3-table3.sh}{}
... \reproduce{./3-fig1.sh}{0}
\end{verbatim}

(Here the fig generating script has up to two arguments.)  The manuscript.pdf file can then be generated by the standard Latex command

\begin{verbatim}
pdflatex manuscript.tex
\end{verbatim}

and the embedded macros will expand within the published manuscript itself for the above example as \textit{``The reproducibility script is \textbf{./3-table3.sh}''} and \textit{``The reproducibility script is \textbf{./3-fig1.sh 0}''} respectively, providing a cross-reference between each claim and the full means to reproduce it.

Similarly, all the necessary scripts to be run can be generated from the same Latex manuscript source file by the rather more eye-watering

\begin{verbatim}
cat InformationInProteomes.tex \
 | perl -ne \
   '/reproduce{([^}]+)}{([0-9]*)}/ && printf "%s %s\n", $1, $2;' \
 | sort
\end{verbatim}

This extracts the necessary scripts in the right order from the .tex file.  In a scientific paper, the ability to repeat the figures and tables one by one like this seems completely fundamental; every diagram and table is accompanied by the complete means to reproduce it locally.  The methods for doing this can be inspected internally throughout the flow, including the statistical analyses, but the primary output is the diagrams in encapsulated postscript form, (.eps).  Assuming the unadulterated data only is being processed, (option 0 on the figure command line generators),

\begin{itemize}
 \item Command-line: \textbf{./3-fig1.sh 0}
 \item Command-line: \textbf{./3-fig2.sh 0}
 \item Command-line: \textbf{./3-fig3.sh 0}
 \item Command-line: \textbf{./3-fig4.sh 0}
 \item Command-line: \textbf{./3-fig5.sh 0}
 \item Command-line: \textbf{./3-fig6.sh 0}
 \item Command-line: \textbf{./3-fig7.sh 0}
 \item Command-line: \textbf{./3-table3.sh}
 \item Command-line: \textbf{./3-table5.sh}
\end{itemize}

Each script and its supporting scripts (which are all included in the deliverability package), then produces the individual components of each figure, sub-figure and table which involved computation.  For example Figure 1 comprised Figures 1A, 1B and 1C.  Each is produced individually. so three .eps files result.

Figures 2, 3 and 4 each include 3-D plots produced using \textit{R}.  The \textit{R} graphics scripts to generate these are also included, so the output of the above scripts on a compliant machine will be exactly as reproduced in the original article in PLOS ONE \cite{HattonWarr2015}.

Finally the 3-table3,5.sh scripts collect the data and apply statistical analysis using the \textit{R} scripts provided.

We note in passing that the above process is significantly more difficult to achieve on proprietary systems such as Windows although perfectly possible \cite{Mesirov2009}, albeit without the complete transparency available in fully open source systems.  In particular we re-iterate the dangers of \textit{any} proprietary data formats.

\subsection*{Verification}
No reproducibility method would be complete without the means to regress the locally produced results with those produced in the published paper, (i.e. the benchmark versions).  Regression is a testing method whereby a freshly generated set of test results is compared with some Gold standard of test results, for example the published and packaged results, and differences quantified and investigated.  To that end, all the generated data (.dat) files used to create the figures and tables can be verified against the benchmark versions included in a sub-directory of the reproducibility package.  The appropriate command in this case is

\begin{itemize}
 \item Command-line: \textbf{./4-verification.sh}
\end{itemize}

This script merely automates the versatile Linux \textit{diff} program as follows

\begin{verbatim}
diff -yw $gold_standard.dat $locally_produced.dat    > differences.dat
\end{verbatim} 

Any differences are then logged in the differences.dat file and can be inspected.  In cases where exact agreement is not expected, for example comparing results run on a 32-bit machine with those on a 64-bit machine, the differences can be quantified by piping the output through a simple program (hypothetical here as we compared integer results in our study and did not need this) as below:

\begin{verbatim}
diff -yw $gold_standard.dat $locally_produced.dat    \
  | quantify_differences.pl > differences.dat
\end{verbatim}

By this means, all locally generated data files can be quickly checked for agreement with the benchmark version.  We note in passing that it is easily possible to verify any .eps graphics files in the same way as .eps files are human readable although their format is not exactly transparent to the casual viewer, but it is harder to quantify any differences and it is probably sufficient to ensure that the data used to draw the final figures agrees with the benchmark version.

It goes without saying that this is not verifying correctness of research; it only verifies the reproducibility of the research, \cite{Peng2011}.

\section*{Conclusion}
Generating fully reproducible research whereby both a manuscript and all the figures and tables can be produced automatically from a single manuscript file is entirely feasible as described here.  It is painstaking but not particularly difficult work but developments in modern open source at least make it possible.  Even though it takes careful preparation, it is greatly comforting to take a single manuscript file as published and from it generate everything, diagram by diagram, statistical analysis by statistical analysis, regress them against a master copy and get the answers as published.

It may be that scientists in other disciplines will need technical and financial resources to achieve this and this should be allowed in the funding, but the real question if we are to take reproducibility and the scientific method seriously amidst the unprecedented growth of computation in science, is can we afford not to do this ?

\section*{Acknowledgements}

\begin{itemize}
 \item We would like to thank all the open source contributors to Linux, \textit{perl} and \textit{R} who in the last 20 years, along with the foresight of the biological community in providing open access to the biological data itself in a compliant form, have allowed computationally intensive biological research to be taken out of the lab and straight onto any reasonable desktop where it can be independently reproduced using similar principles to those espoused here.  These are opportunities of which scientists as a whole should make the most.
 \item \textbf{Competing Interests:} The authors declare that they have no
competing financial interests.
 \item \textbf{Correspondence:} Correspondence and requests for materials
should be addressed to Les Hatton ~(email: lesh@oakcomp.co.uk).
 \item This material is based in part on work supported by the National Science Foundation. Any opinion, finding, and conclusions or recommendations expressed in this material are those of the author and do not necessarily reflect the views of the National Science Foundation.
\end{itemize}

\nolinenumbers

%
%
%
\bibliography{lh_biblio}

\end{document}